\documentclass[a4paper]{jpconf}
\usepackage{graphicx}
\usepackage[english,russian]{babel}
\usepackage{amsmath,amssymb}

\newcommand{\wl}{{\omega_{\ell}}}
\newcommand{\al}{\alpha}
\usepackage{euscript}   

\newcommand{\be}{\begin{equation}}
\newcommand{\ee}{\end{equation}}

\newcommand{\comment}[1]{}  

\begin{document}
\selectlanguage{english}
\title{Modeling of the magnetic properties of~nanomaterials with different crystalline structure }

\author{Yury Kirienko and Leonid Afremov}

\address{Department of Theoretical and Experimental Physics,
School of Natural Sciences, Far-Eastern Federal University, 8, Sukhanova str.,
Vladivostok, Russia}

\ead{yury.kirienko@gmail.com}

\begin{abstract}
We propose a method for modeling the magnetic properties of nanomaterials with different structures. 
The method is based on the Ising model and the approximation of the random field interaction. 
It is shown that in this approximation, the magnetization of the nanocrystal depends only 
on the number of nearest neighbors of the lattice atoms and the values of exchange integrals between them. 
This gives a good algorithmic problem of calculating the magnetization of any nano-object, 
whether it is ultrathin film or nanoparticle of any shape and structure, 
managing only a rule of selection of nearest neighbors. 
By setting different values of exchange integrals, it is easy to describe ferromagnets, 
antiferromagnets, and ferrimagnets in a unified formalism. 
Having obtained the magnetization curve of the sample it is possible to find the Curie temperature 
as a function of, for example, the thickness of ultrathin film. 
Afterwards one can obtain the numerical values for critical exponents of the phase transition ``ferromagnet -- paramagnet". 
\comment{%
It is shown that the obtained values do not depend on the type of crystal lattice, 
and the numerical values coincide with the values obtained by the method of renormalization group.
}
Good agreement between the results of calculations and the experimental data
proves the correctness of the method.
\end{abstract}

\section{Introduction}
Creation of magnetic materials with predetermined properties -- the task of unquestionable importance.
While the manipulation of individual atoms becomes a routine, 
nanotechnologies need appropriate calculation methods, 
which would allow prototyping magnetic properties of materials without heavy calculations on supercomputers.
The method proposed below satisfies such requirements.
Moreover, it allows calculus of the properties of magnets of any type 
(ferromagnets, antiferromagnets, ferrimagnets) to be provided using a uniform formalism.

Another feature of nano-objects is the existence of size effects at their scale.
Obviously, such effects are different in two-dimensional films and three-dimensional particles.
Furthermore, it is common to use different methods of modeling for nanoparticles of different shape.
But if to consider the theory with short-range interaction, the global geometry of the material should not play any role.
In this case, there must be a simulation method that is insensitive to the shape of the simulated sample as a whole.
We offer in this paper such a method.

\section{General model of magnetic material}
We assume that magnetic atoms are distributed over $N$ sites of the sample with probability $p$. 
According to~\cite{Belokon2001}, the distribution function 
for random interaction fields $H$ on a particle located at the origin can be defined as:
\be\label{eq.1}
W\left(H\right)=
\int{\delta \left(H-\sum_k{h_k\left({{\mathbf r}}_k,{{\mathbf m}}_k\right)}
\prod_k{F\left({{\mathbf m}}_k\right)\delta \left({{\mathbf r}}_k-{{\mathbf r}}_{k0}\right)}\right)d{{\mathbf r}}_kd{{\mathbf m}}_k},
\ee
where $\delta(x-x_0)$~-- Dirac delta function, $h_k=h_k\left({{\mathbf m}}_k,{{\mathbf r}}_k\right)$~-- 
field created by atoms with magnetic moments ${{\mathbf m}}_k$ located at coordinates ${{\mathbf r}}_k$,
${{\mathbf r}}_{k0}$~-- the coordinates of lattice sites, $F({{\mathbf m}}_k)$~-- the~distribution 
function for the magnetic moments, which in the approximation of Ising model for a ferromagnets can be 
represented as follows:
\be\label{eq.2}
F\left({{\mathbf m}}_k\right)=
\left({\alpha }_k\delta \left({\theta }_k\right)+{\beta }_k\delta \left({\theta }_k-\pi \right)\right)
\left(\left(1-p\right)\delta \left(m_k\right)+p\delta \left(m_k-m_0\right)\right).
\ee
Here ${\theta}_k$~-- the angle between ${{\mathbf m}}_k$ and $OZ$-axis, $\alpha_k$ and $\beta_k$~--
relative probabilities of the spin orientation along and against $OZ$-axis 
(${\theta}_k=0$ and ${\theta}_k=\pi$, respectively); 
$m_0$ -- magnitude of magnetic moment of a magnetic atom. 
Probabilities $\al_k$ and $\beta_k$ hold normalization condition $\al_k+\beta_k=1$.
In the approximation of nearest neighbors and the direct exchange interaction between magnetic atoms, 
the equation \eqref{eq.1} can be represented as:
\be\label{eq.3}
W_j\left(H\right)=
\sum^z_{n=0}{p^{z-n}{\left(1-p\right)}^n\sum^{C^{z-n}_z(k_j)}_{\nu }
{\sum^{2^n}_{l_{\nu }\in L\left(C^n_z\left(k_j\right)\right)}{{\omega }_{l_{\nu }}\delta \left(H-M_{l_{\nu }}J_j\right)}}},
\ee
where $k_j$ is a set of nearest neighbors of the magnetic atom numbered as $j$, 
$z={\dim k_j}$ -- its coordination number; 
$C^n_z\left(k_j\right)$ -- subset of $n$ atoms of the total number of $z$ nearest neighbors of $j^{th}$ atom; 
$L\left(\Omega \right)$ is a binomial set of permutations of an arbitrary set $\Omega $ with the amount of elements equal to $2^{{\dim  \Omega}}$.
Introducing symmetric notation ${\alpha }_{-n}\equiv1-{\alpha}_n$ we have got 
${\omega}_{l_j}=
\prod^{\nu=z}_{(\nu =1;l_{\nu }\in k_j)}
{{{\alpha }_{\pm l}}_v}$ and $M_{l_j}=\sum_{n\in k_j}{\pm m_n}=m_0\sum_{n\in k_j}{\pm |{2\alpha }_n-1|}$. 
Finally, $J_j$ is the constant of exchange interaction (exchange integral).

Using the expression for the distribution function of interaction fields \eqref{eq.3}, 
one can obtain equations that determine average relative magnetic moments at each site of lattice:
\be\label{eq.4}
{\mu }_j=\int{{\tanh\left(\frac{m_jH}{k_BT}\right)W_{j}\left(H\right)dH
        =\sum^z_{n=0}{p^{z-n}{\left(1-p\right)}^n\sum^{C^{z-n}_z(k_j)}_{\nu }
            {\sum^{2^n}_{l_{\nu}\in L\left(C^n_z\left(k_j\right)\right)}
            {{\omega}_{l_{\nu }}{\tanh  \left(\frac{M_{l_{\nu }}J_j}{k_BT}\right)}}}}.}}
\ee
Expression~\eqref{eq.4} allows to investigate the dependence of total magnetic moment of the sample $M=\sum^{N}_{j=1}{{\mu}_j}$ 
at the temperature $T$ and concentration $p$, as well as to determine the dependence on the number $N$ of atoms 
of the temperature of phase transition
and the percolation threshold. 
System of $N$ equations with $N$ unknowns~\eqref{eq.4} can be solved numerically using Newton's method\footnote{
In our research we use {\tt Wolfram Mathematica} for fast lattice prototyping and {\tt C++} and {\tt Fortran} non-linear solvers 
with {\tt python} wrapper for precise calculus.}.

\section{Modeling of magnetic materials with different crystalline structure}

Equation~\eqref{eq.4}, despite the complicated form, has written in the algorithmically convenient form.
This form allows to simulate the magnetic properties of materials, 
based only on the knowledge of the crystalline structure of the sample
and the numerical values of the exchange integrals. 

Crucial part of this method is the rule of selection of nearest neighbors for each atom
(the way of constructing of the set $k_j$). By changing this rule of selection we can easily
adjust our model to different physical systems. 
We also can simulate antiferromagnetic and ferrimagnetic materials by
reversing signs and values of exchange integrals of individual atoms.

Consider two examples of applying the method described above.

\subsection{Ultrafine particles}
Consider a nanoparticle of $N$ atoms. 
Then~\eqref{eq.4} is a system of $N$ independent equations with $N$ unknowns.
In some cases, when $p=1$, the symmetry of the particle reduces the number of unknowns,
but in general all variables are different.

For example, for cubic-shaped nanoprticle with $n$ atoms on the edge and simple cubic lattice
there is a system of $n^3$ non-algebraic equations with $n^3$ unknowns.
Even relatively small cubic particle with $10$ atoms on the edge contains $10^3$ atoms,
which makes calculation non-trivial and significantly reduces its accuracy.

\subsection{Thin films}
\begin{figure}
    \includegraphics[width=0.45\textwidth]{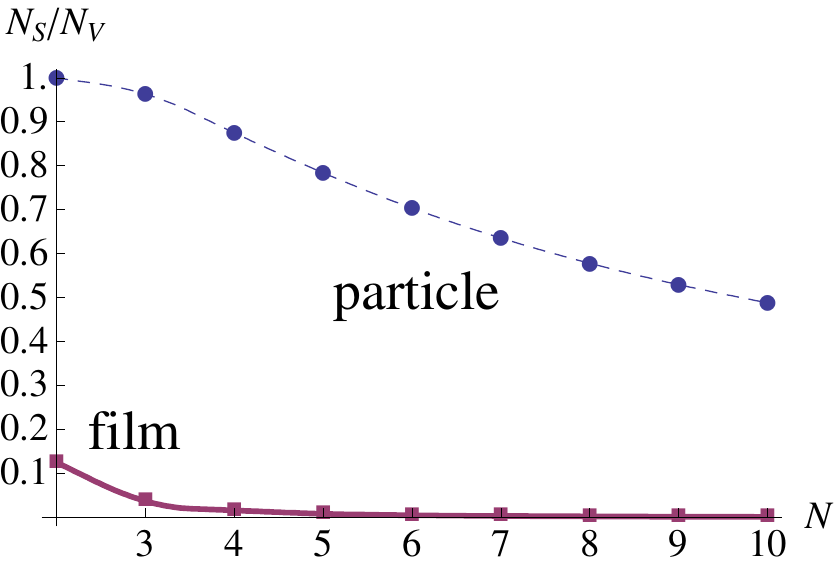}\hspace{0.1\textwidth}%
\begin{minipage}[b]{0.45\textwidth}\caption{\label{fig.1}%
    The ratio of the number of atoms on the surface of the sample $N_S$ 
    to the number of atoms in the volume $N_V$ as a function of size 
    (for cubic particles $N$ is the number of atoms on the edge, for films~-- its thickness).}
\end{minipage}
\end{figure}

Now consider ultrathin film that is composed of $N$ infinite monolayers.
In a simple case when all atoms from the same layer are equal, 
\eqref{eq.4} turns into a system of $N$ equations with $N$ unknowns.
Study of size effects in films is much easier than in the particles, 
because it is sufficient to simulate a film of 10-15 layers
to achieve the bulk properties.
(See fig.~\ref{fig.1}.)

\begin{figure}
    \includegraphics[width=0.5\textwidth]{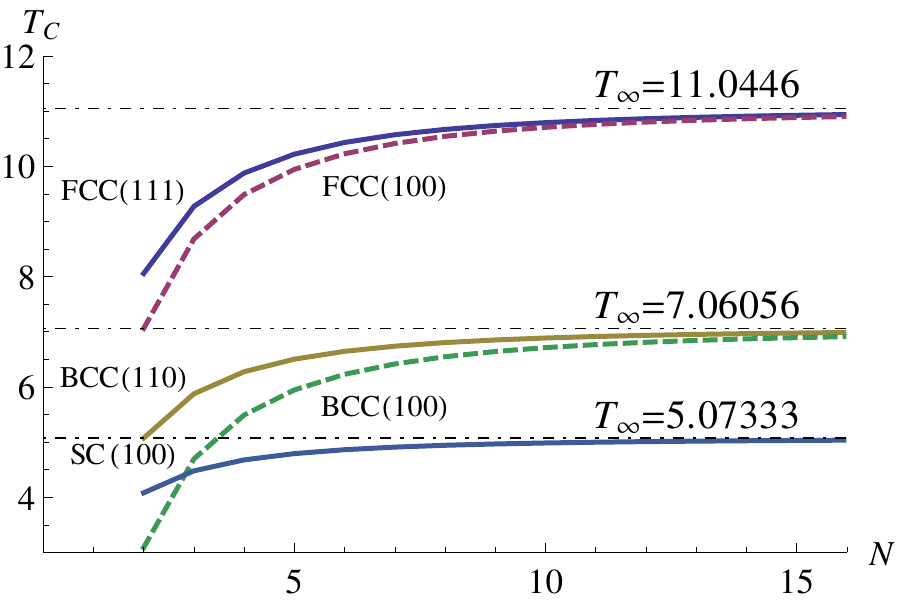}\hspace{0.1\textwidth}%
\begin{minipage}[b]{0.4\textwidth}\caption{\label{fig.2}%
    The dependence of reduced Curie temperature $T_C$ on thickness of ultrathin film 
    ($N$ is the number of monolayers) for different crystal lattices (FCC,  BCC and SC) and different 
    crystallographic orientations of the surface. }
\end{minipage}
\end{figure}

General equation that determines the average relative magnetic moment ${\mu}_n$ in $n$-th monolayer is
given by~\eqref{eq.4}.
Replacing in expressions for $\wl$ and $M_{\ell}$ all ${\al}_{\pm n}$ on their average values 
$\left\langle {\al}_{\pm n}\right\rangle ={(1\pm {\mu}_n)}/{2}$ 
and substituting \eqref{eq.3} into \eqref{eq.4}, 
one can obtain the equations that determine ${\mu}_n$ in each monolayer:
\be  \label{eq.5} 
\left\{
\begin{array}{rl}
    \mu_1&=
        \sum\limits^{z_{1,1}}_{l=0}{\left(l\atop{z_{1,1}}\right)}
            {\langle\alpha_1\rangle}^l{\langle\beta_1\rangle}^{z_{1,1}-l}
        \sum\limits^{z_{1,2}}_{k=0}{\left(k\atop{z_{1,2}}\right)}
            {\langle\al_2\rangle}^k{\langle\beta_2\rangle}^{z_{1,2}-k}
        {\tanh \left(\frac{\left(2l-z_{1,1}\right)+(2k-z_{1,2})i_{1,2}}{t}\right)\ },\\
    \mu_n &=
        \sum\limits^{z_{n,n}}_{l=0}{\left(l\atop{z_{n,n}}\right)}
            {\langle\alpha_n\rangle}^l{\langle\beta_n\rangle}^{z_{n,n}-l} 
        \sum\limits^{z_{n-1,n}}_{k=0}{\left(k\atop{z_{n-1,n}}\right)}
            {\langle\alpha_{n-1}\rangle}^k{\langle\beta_{n-1}\rangle}^{z_{n-1,n}-k}\times\\ 
        &\qquad \sum\limits^{z_{n,n+1}}_{r=0}
            {\left(r\atop{z_{n,n+1}}\right)}
                {\langle\alpha_{n+1}\rangle}^r{\langle\beta_{n+1}\rangle}^{z_{n,n+1}-r}\times\\ 
        &\qquad \qquad
             \tanh  \left(\frac{\left(2k-z_{n-1,n}\right)i_{n-1,n}
                                  +\left(2l-z_{n,n}\right)i_{n,n}
                                  +\left(2r-z_{n,n+1}\right)i_{n,n+1})}{t}\right)\\
    \mu_N &=
        \sum\limits^{z_{N,N}}_{l=0}{\left(l\atop{z_{N,N}}\right)}
            {\langle\alpha_N\rangle}^l{\langle\beta_N\rangle}^{z_{N,N}-l}
        \sum\limits^{z_{N-1,N}}_{k=0}{\left(k\atop{z_{N-1,N}}\right)}
            {\langle\al_{N-1}\rangle}^k{\langle\beta_{N-1}\rangle}^{z_{N-1,N}-k} \times\\
        &\qquad \tanh \left(\frac{\left(2l-z_{N-1,N}\right)i_{N-1,N}+(2k-z_{N,N})i_{N,N}}{t}\right),   
\end{array}
\right.
\ee 
where $z_{n,n}$ is the number of nearest neighbors in the $n$-th layer,
$z_{n-1,n}$ is the number of nearest neighbors of the atom in $(n-1)$-th layer, located  in the $n$-th layer;
$i_{nn}={J_{nn}m_n}/{J_{11}}m_1$, $i_{n-1,n}={J_{n-1,n}m_{n-1}}/{J_{11}m_1}$, $i_{n,n+1}={J_{n,n+1}m_{n+1}}/{J_{11}m_1}$, 
$t={k\ T}/{J_{11}}m_1$. Using  \eqref{eq.5}, one can study the dependence of 
the average magnetic moment of the film on its temperature and thickness~\cite{Hongkong2012},
and, as a consequence, the dependence of Curie temperature on the thickness (see fig.~\ref{fig.2}).

\section{Comparison with experimental data}
Comparison of the results of modeling with experimental data was carried out directly and indirectly.
\subsection{Direct comparison: relative Curie temperature change}
In~\cite{Xiamen2012}, the dependence of Curie temperature $T_C(N)$ on the size $N$ 
of cubic-shaped particle was calculated. 
And then the magnitude of the relative change of the Curie temperature 
$\varepsilon(N)=T_C(N)/T_C(N\to\infty)$ was obtained. 
Comparison with experimental data~\cite{Sadeh2000} revealed the validity of the modeling method.
\subsection{Indirect comparison: critical exponents}
It is possible to obtain the value of critical exponent $\nu$ of spin-spin correlation 
for phase transition ``ferromagnet~-- paramagnet'' 
by appropriate approximation of $\varepsilon(N)$ defined above. Critical exponents for ultrathin films
of different crystalline structures were obtained in~\cite{Hongkong2012}. It was shown that the value of $\nu$
for three-dimensional Ising model is independent from the type of the lattice. The numerical value of $\nu$ is 
close to the value obtained from RG-calculations~\cite{Guillou1977}.
\section{Conclusion}
Offered method of simulating allows to substitute the real magnetic material with the relatively simple model, 
where the key role is played by the coordination number and the exchange integrals. 
All the magnetic geometry lies in these two integral characteristics. 
By setting only the rules of choice of nearest neighbors and values of corresponding 
exchange integrals, we make the modeling procedure very effective.

\ack
\selectlanguage{russian}
The work was supported by grant of 
Scientific Fund of Far Eastern Federal University (FEFU)
\No\,12-07-13000-FEFU\_a.
\selectlanguage{english}

\section*{References}
\bibliographystyle{iopart-num}
\bibliography{hungary}

\providecommand{\newblock}{}
\begin{thebibliography}{1}
\expandafter\ifx\csname url\endcsname\relax
  \def\url#1{{\tt #1}}\fi
\expandafter\ifx\csname urlprefix\endcsname\relax\def\urlprefix{URL }\fi
\providecommand{\eprint}[2][]{\url{#2}}

\bibitem{Belokon2001}
Belokon V and Nefedev K 2001 {\em Journal of Experimental and Theoretical
  Physics\/} {\bf 93}(1) 136--142 ISSN 1063-7761 10.1134/1.1391530
  \urlprefix\url{http://dx.doi.org/10.1134/1.1391530}

\bibitem{Hongkong2012}
Afremov L and Kirienko Y 2012 {\em Advanced Materials Research\/} {\bf
  378--379} 589--592 (\textit{Preprint}
  \eprint{http://arxiv.org/abs/1108.0745})

\bibitem{Xiamen2012}
Kirienko Y and Afremov L 2012 {\em Advanced Materials Research\/} {\bf
  472--473} 1827--1830 (\textit{Preprint}
  \eprint{http://arxiv.org/abs/1201.1562})

\bibitem{Sadeh2000}
Sadeh B, Doi M, Shimizu T and Matsui M 2000 {\em Journal of the Magnetics
  Society of Japan\/} {\bf 24} 511--514

\bibitem{Guillou1977}
Le~Guillou J and Zinn-Justin J 1977 {\em Phys. Rev. Lett.\/} {\bf 39} 95--98

\end{thebibliography}
\end{document}